\documentclass{ecai}
\usepackage{times}
\usepackage{graphicx}
\usepackage{latexsym}
\usepackage[hyphens]{url}
\usepackage{xcolor}

\newcommand{\quotes}[1]{``#1''}

\newcount\Comments  
\newcommand{\kibitz}[2]{\ifnum\Comments=1{\textcolor{#1}{#2}}\fi}

\begin{document}

\title{Applying Transparency in Artificial Intelligence based Personalization Systems} 


\author{Laura Schelenz\footnote{University of Tübingen,
Germany, email: laura.schelenz@uni-tuebingen.de}
\and Avi Segal\footnote{Ben-Gurion University of the Negev,
Israel, email: avisegal@gmail.com} \and Kobi Gal\footnote{The University of Edinburgh, UK, email: kgal@inf.ed.ac.uk}}

\maketitle
\bibliographystyle{ecai}

\begin{abstract}


Artificial Intelligence based systems increasingly use personalization to provide users with relevant content, products, and solutions. Personalization is intended to support users and address their respective needs and preferences. However,  users are becoming  increasingly vulnerable to online manipulation due to algorithmic advancements and lack of transparency. Such manipulation decreases users’ levels of trust, autonomy, and satisfaction concerning the systems with which they interact. Increasing transparency is an important goal for personalization based systems. Unfortunately, system designers lack guidance in assessing and implementing transparency in their developed systems.

In this work we combine insights from technology ethics and computer science to generate a list of transparency best practices for machine generated personalization. Based on these best practices, we develop a checklist to be used by designers wishing to evaluate and increase the transparency of their algorithmic systems. Adopting a designer perspective, we  apply the checklist to prominent online services and discuss its advantages and shortcomings. We encourage researchers to adopt the checklist in various environments and to work towards a consensus-based tool for measuring transparency in the
personalization community.


\end{abstract}

\section{Introduction}
Recent years saw significant increase in personalization approaches for online systems~\cite{segal2018optimizing,van2019collecting,yanovsky2019one}. Such personalization can be  used to direct users' attention to relevant content~\cite{kolekar2019rule}, increase their motivation when working online~\cite{muldoon2018survey}, improve their performance~\cite{jackson2016encouraging}, extend their engagement~\cite{segal2016intervention} and more. These approaches rely on social theories of human behaviour (e.g.~\cite{deci2002overview,borji2012state}) as well as on machine learning based abilities  to predict human behavior and human response to various interventions~\cite{plonsky2019predicting,hartford2016deep}. 

Yet, personalization technology that focuses on maximizing system designers goals runs the risk of marginalizing users~\cite{link_uxdesign}.
Personalized recommendations of content, products, and services usually attempt to influence a person's decision-making. When such influences are hidden and subtly try to persuade users (maybe even against their expressed goals), this constitutes a form of manipulation~\cite{Susser.2018}. Subverting a person's decision-making abilities  reduces  their autonomy. Especially with regard to personalized advertisement, personlization can  exploit users' vulnerabilities~\cite{Spencer.2019} and may even threaten democratic processes~\cite{Crain.2019}.
Applying transparency to the design of personalized content can help address the above mentioned challenges. Indeed, the computer science community has affirmed the importance of transparency in its profession. The ACM Code of Ethics reads: \quotes{The entire computing profession benefits when the ethical decision-making process is accountable to and transparent to all stakeholders}~\cite{ACMCouncil.22June2018}. Political bodies also identify transparency as a pivotal principle in Artificial Intelligence based software~\cite{Floridi.2018, HighLevelExpertGrouponArtificialIntelligence.2019}. Especially with the advent of legal frameworks that prescribe transparency in data collection, processing, and storage~\cite{EuropeanParliamentandCounciloftheEuropeanUnion.27April2016}, system designers require increased awareness and guidance in the implementation of transparency in their systems. Recent efforts recognize this need for guidance and seek to synthesize and make ethics principles more tangible for implementation in systems~\cite{Amershi.2019, AIEthicsImpactGroup.2020}.

We begin our analysis by discussing the need for transparency in Artificial Intelligence systems (section~\ref{sec:need}). 
We then provide a definition of transparency drawing on prior art and address transparency's distinction from explainability and related concepts (section~\ref{section:transtopics}). We identify transparency as a practice of communication and interaction with the end user regarding a system's features and possible effects, including options for user control.
We move on to derive transparency best practices for personalization in online systems (section~\ref{section:bestpractices}). These best practices constitute ethical responsibilities on the part of system designers. Based on these best practices, we specify questions that should be addressed when considering transparency in personalization processes. This constitutes a concrete first checklist that can be used by system designers to evaluate and operationalize transparency in their systems (section~\ref{section:checklist}). We then describe a preliminary application of the checklist to  existing web sites in the wild (section~\ref{section:apply}). Specifically, we look at Facebook, Netflix, YouTube, Spotify and Amazon. For each such destination, we check how the elements from the checklist are supported in the particular site. We finish by discussing the value but also limitations of our approach and by pointing to further research needed in this area (sections~\ref{section:discuss} and~\ref{section:conclude}).

We make the following main contributions: we create a new definition of transparency in the context of AI based personalization; we develop a set of best practices for the community based on this definition and on prior art; we generate a concrete tool-set to help system designers asses and realize these best practices in their respective systems.
 
\section{The Need for Transparency in AI Based Systems}
\label{sec:need}
The speedy uptake of Artifical Intelligence based approaches  has raised concerns about their ambivalence. Personalized recommendation may help users find relevant content and items but also introduce bias and pose a risk of manipulation~\cite{Bozdag.2013}. Similarly, algorithmic decision-making may help allocate resources in a fairer manner but also pose risks of discriminating against social groups due to bias in the system~\cite{Zou.2018}. Such risks are especially high if the user is unaware of the processes of personalization and decision making. This is often the case with algorithmic systems, as filtering, classification, personalization, and recommendation remain intransparent or even opaque~\cite{Pasquale.2015,Burrell.2016}.

Transparency can help address the concerns voiced about AI based personalization systems. First, transparency can balance power asymmetry, empowering users while curtailing the influence of companies on customer behavior. While companies have easy access to user data, users lack knowledge of algorithmic systems~\cite{Lepri.2018}. Especially big players in the information system economy hold enormous power vis à vis users to the extent that they can shape information, knowledge and culture production~\cite{Benkler.2006}. User empowerment by means of transparency and user control may level the playing field. 

Second, transparency can increase user autonomy. Recommender systems usually filter content according to preference models that easily create a feedback loop. A classic example is the “filter bubble” in social media platforms~\cite{Pariser.2012,Veltri.2019}. When   users lack  exposure to information diversity, their autonomy and ability to make independent decisions is impacted~\cite{Mittelstadt.2016}. However, if users understand why and how an algorithm presents information to them, they can better reflect on how sources of information inform their decisions.

Third, transparency can boost privacy rights and user trust in algorithmic systems. Users can only give meaningful informed consent when they understand the risks of algorithmic decision-making~\cite{Mittelstadt.2016}. 

Fourth, transparency can enable fairness and non-discrimination in algorithmic decision-making. Algorithmic decision-making is becoming ever more pervasive, affecting individuals in pivotal areas of life~\cite{Laat.2018}. While human decision-making reserves the possibility to provide a straightforward face to face explanation of why someone’s application was denied, algorithmic systems are considered too complex for operators to provide a simple answer~\cite{Miller.22062017} (for an opposing view, see~\cite{Zerilli.2019}). Transparency may thus increase subjects’ ability to understand the cause of decisions made by algorithms. Thereby, transparency enables users to assess whether a decision-making process is fair and non-discriminatory~\cite{Abdollahi.C2018}. 

While transparency is highly relevant, it is not absolute. Calls for transparency may not always be ethical and warranted. Indeed, they depend on the standing of different actors that are involved and interact in algorithmic assemblages~\cite{Kitchin.2017,Seaver.2019,Brill.2015}. For instance, demanding increased transparency on behalf of users (in terms of sharing more data) seems inappropriate given their vulnerability to loss of informational privacy~\cite{Lanzing.2016}. It is thus appropriate to focus attention on promoting transparency  from the system design perspective, and increase users understanding of the logic underlying designers' activities~\cite{Kitchin.2017,ONeil.2016,Steiner.2013}.

\section{Best Practices for Transparency} 
\label{section:transdef}
We take a three step approach to developing best practices for transparency in machine generated personalization. First, we develop a new definition of transparency for algorithmic systems by drawing on prior art. Second, from this definition, we derive best practice for the implementation of transparency in machine generated personalization. Third, we translate these best practices into questions for system designers to be used as a reflection and assessment tool, presented as an online checklist for open usage. 


\subsection{Step 1: Transparency Definition}
\label{section:transtopics}

To generate a list of best practices, we began by asking: What is
transparency in the context of AI systems? When working with the
term transparency, we should first clarify the relationship of transparency to principles of ethics. According to Turilli and Floridi~\cite{Turilli.2009}, transparency is not an ethics principles itself. Rather, transparency can enable or prevent ethics. In some cases, calls for transparency may for instance inhibit privacy rights. We thus frame transparency not as a principle of ethics but as a practice that can achieve ethics goals such as autonomy and accountability~\cite{Turilli.2009}.
We investigated views on transparency from technology ethics, the
philosophy of technology, computer sciences, but also ethics guidelines and legal documents. Based on our analysis, we define transparency
as follows:

\emph{Transparency is a practice of system design that centers on the disclosure of information to users, whereas this information should be understandable to the respective user and provide insights about the system. Specifically, the information disclosed should enable the user to understand why and how the system may produce or why and how it has produced a certain outcome (e.g. why a user received a certain personalized recommendation).}

Recent and emerging scholarship on explainable AI has provided computational methods to increase explainability in computer systems~\cite{Hois.2019, Rosenfeld.2019, Zhou.C2018, Preece.2018, adadi2018peeking,unknown}. While transparency is often used synonymously with explainability and similar concepts such as observability, controllability, and interpretibility, transparency is in fact broader than  explaining a system's functionality or enabling the user to infer information from a system's outcome. Transparency follows a more comprehensive approach, as it combines several components, which we now lay out.

The first important component of transparency is the notion that information must be \quotes{understandable}. The user of a system must be able to comprehend the information disclosed to them. For instance, the GDPR~\cite{EuropeanParliamentandCounciloftheEuropeanUnion.27April2016} states with regard to data processing that information must be provided in \quotes{clear and plain language and it should not contain unfair terms}~\cite{EuropeanParliamentandCounciloftheEuropeanUnion.27April2016}. 

Here, we can see how transparency is a relational concept and a performative practice~\cite{Ananny.2018}. Whether the information provided is transparent to an individual user (or data subject) depends on their cognitive abilities, their language skills, and epistemic conditions. Therefore, practices of transparency must be personalized to the user at hand, given the diversity of users’ ability to comprehend information~\cite{Vakarelov.2019}. 

Several sources stress the importance of information comprehensibility. According to Chromnik et al.~\cite{Chromnik.20March2019}, transparency is an enabling condition for the user to \quotes{understand the cause of a decision}. Ananny and Crawford~\cite{Ananny.2018} describe transparency as a form of “seeing” and “understanding” an actor-network. The authors stress that transparency means not merely looking inside a system but across systems. Transparency thus means explaining a model as it interacts with other actors in an algorithmic system~\cite{Ananny.2018}. Floridi et al.~\cite{Floridi.2018} understand transparency as explainability, whereas explainability incorporates both intelligibility and accountability.
AI decision-making processes can only be understood if we are able to grasp how models work and who is responsible for the way they work~\cite{Floridi.2018}. For Vakarelov and Rogerson~\cite{Vakarelov.2019}, transparency means communication of information under two conditions: information must be a) sufficient and b) accessible. The latter means that the recipient of the information must be able to comprehend and act upon the information. 

Another crucial element of transparency is information disclosure about deliberation or decision-making processes. The IEEE Guideline for “Ethically Aligned Design” states that transparency means the possibility to ascertain why a certain decision was made~\cite{InstituteofElectricalandElectronicsEngineers.2019}. For Turilli and Floridi~\cite{Turilli.2009}, disclosing information refers to communication about the deliberation process, i.e. how information came about. The rationale here is that the deliberation process reveals the values that guide organizations or system designers in their everyday practices and illustrate how they make decisions. 

Similarly, for Tene and Polonetsky~\cite{Tene.2013}, transparency refers to the revelation of information about criteria used in decision-making processes. The disclosure of the dataset (or its existence) is less relevant than the actual factors (such as inferences made from the data) that inform a model and its effects on users. Also Zerilli et al.~\cite{Zerilli.2019} argue that, similar to explanations in human decision-making, a system should reveal factors in decision-making and how they might be weighted. 

Dahl~\cite{Dahl.2018} even argues that it is not necessary to reveal the inner working of a model for the user to determine whether a system is trustworthy. Rather, transparency means providing key details about how the results came about or offering expert testimony about how the system usually works. Burrell~\cite{Burrell.2016} suggests that improving interpretability of models is crucial to reduce opacity: “One approach to building more interpretable classifiers is to implement an end user facing component to provide not only the classification outcome, but also exposing some of the logic of this classification.” 

Finally, there can be an element of participation in transparency. The user is expected to assess the system with regard to its trustworthiness based on the information that is disclosed. Furthermore, the user may become active in choosing between different models, i.e. different options of personalization~\cite{Simon.2010}. The user is thus becoming involved in the process of transparency which increases user control while interacting with the system.

\subsection{Step 2: Best Practices}
\label{section:bestpractices}
From our definition of transparency, we derived nine principles of transparency for responsible personalization. 
They reflect the three core elements of transparency: information provided must be understandable to users, information must be disclosed about why and how a model reaches a certain outcome, and users should have a say in personalization processes. The best practices further reflect additional needs for information about the data collection processes, the composition of datasets, the functionality of a model, the responsibilities for the model or system, and how the model may interact with other models across algorithmic systems. 

Table~\ref{tab:dim} shows the list of the best practices as well as the sources on which these practices build. Based on the qualitative analysis in step 1, particular relevance can be ascribed to practices 1, 2, 3, and 8. The table also identifies the different system architecture components relevant for each best practice based on the Input-Processing-Output architecture model~\cite{boell2012conceptualizing}. These components include: \quotes{Input} for transparency relating to the data used by the system, \quotes{Processing} for transparency relating to system models and \quotes{Output} for presenting the transparent information to the user. We extend this architecture with a \quotes{Control} component to represent the control given to the user over the system's personalization behaviour. 

We define user control as the possibility of users to interact with the system to adjust elements thereof to their respective needs and preferences. It is important that users not only “feel” that they have control because this can put them at risk of exploitation. If users think that they have control, they might feel encouraged to share more data~\cite{Tene.2013}. User control is thus of particular ethical sensitivity and significance as it relates directly to the autonomy of a person. Past research has demonstrated the importance of user control mechanisms in Artificial Intelligence based systems~\cite{harambam2019designing}.

\begin{table*}[t]
  \centering
 \begin{tabular}{|l|p{2.5cm}|p{7cm}|l|}
\hline

\textbf{No.} & \textbf{Component} & \textbf{Description of transparency standard} & \textbf{Sources} \\
\hline
1 & Input, Processing, Output, Control & Disclosing accessible and actionable information, meaning that the user can comprehend and act upon the information & ~\cite{Vakarelov.2019, EuropeanParliamentandCounciloftheEuropeanUnion.27April2016}  \\
\hline
2 & Input, Processing & Disclosing relevant and detailed information about data collection and processing; this includes notification of data collected for personalization, information about pre-processing and possible biases in the dataset & ~\cite{EuropeanParliamentandCounciloftheEuropeanUnion.27April2016, Buolamwini.2018, ACMCouncil.22June2018}  \\
\hline
3 & Processing & Disclosing relevant and detailed information about the goals of the designer/system, the reasoning of a system, the factors and criteria used (potentially also how they are weighted), as well as the inferences made to reach an algorithmic decision & 
~\cite{Turilli.2009,Floridi.2018, Burrell.2016, Zerilli.2019, Tene.2013, Chromnik.20March2019, Srmo.2005, InstituteofElectricalandElectronicsEngineers.2019, Abdollahi.C2018}
\\
\hline
4 & Processing & Providing expert testimony (e.g. by a member of the design team) about how a system works and reaches a certain outcome, including information about the stochastic nature of a model as well as lab accuracy performance of a model & ~\cite{Dahl.2018} \\
\hline
5 & Processing & Disclosing information about how the algorithmic model may affect the user and how the model may interact with other models across algorithmic systems & ~\cite{Ananny.2018} \\
\hline
6 & Output & Disclosing that a machine is communicating with the user and not a real person & ~\cite{link_latlng_03} \\
\hline
7 & Output & Disclosing information about those responsible for the model (e.g. name of the company or designer) & ~\cite{Floridi.2018}  \\
\hline
8 & Control & Proposing alternative choices for user interaction with the system, e.g. different options for personalization & ~\cite{Simon.2010} \\
\hline
9 & Control & Providing the user with opportunities to give feedback about personalization; providing the user with opportunities to specify their goals as these goals are expected to drive personalization & ~\cite{heer2019agency} \\
\hline
 \end{tabular}
  \caption{Transparency Best Practices for Machine Generated Personalization.}
  \label{tab:dim}
\end{table*}
\subsection{Step 3: Checklist}
\label{section:checklist}
Based on steps 1 and 2, we can now move to define a checklist for systems designers to assess the transparancy of machine generated personalization. We map each architecture component in Table~\ref{tab:dim} (namely Input, Processing, Output, Control) to a section in the checklist. Questions for each section are then derived from the best practices uncovered in the previous steps. In this process, we prioritize some best practices that were overwhelmingly affirmed by the literature.

The resulting checklist is given at:\\ \textbf{\url{http://tiny.cc/evxckz}}.

The checklist includes a total of 23 questions, presented in Table~\ref{tab:app} (described in the next section). After filling it, the system designer can download a PDF file with their responses. They can also print an empty copy of the checklist to be filled offline if needed. 

To arrive at a comprehensible and user friendly checklist, we omitted some questions. If system designers wanted to attempt at particularly high standards of transparency, they could also answer the following additional questions: 
\begin{itemize}
\item Does the system explain to the user how the model(s) may interact with other models in algorithmic systems?

\item Does someone from the design team provide expert testimony to the users about how the model(s) works (e.g. in a video)? 
\end{itemize}



We note that the checklist is supplied as an assessment tool for system designers, enabling them to identify areas in their system which suffer from lack of transparency as well as point to imbalances between the transparency aspects of a system and the control it gives users over its operation. 
Ideally, a system designer has implemented transparency so that they can check yes for every question. However, the goal should not be to score high on the checklist  but rather to have an honest assessment and decide on priorities and next steps.

\section{Case Study: Applying the checklist}
\label{section:apply}
We performed an initial application of the proposed checklist as a reflective and assessment tool for the following online services that use personalization: Facebook, Netflix, YouTube, Spotify, and Amazon. For each of these destinations, we took a system designers point of view, and asked \quotes{how are the transparency elements from the checklist supported on this particular site?}, when examining the information available to registered users on the sites. For this assessment
we adopted the checklist and examined the above web services
using one of the authors account on these sites. Specifically, we checked the information available to registered users on the sites including the privacy policy, the legal terms and conditions and other information that is shared with the user and covers any of the checklist elements. We answered each checklist question for each site with a \quotes{yes}, \quotes{no} or \quotes{partial} reply.

Table~\ref{tab:app} presents our application of the checklist to Facebook. As can be seen from the table, while some transparency elements are well established on this site, other elements are only partially supported or not supported at all and should be considered for future improvement.
\begin{table*}
  \centering
  \scalebox{1.03}{%
 \begin{tabular}{|p{8.5cm}|p{1cm}|p{6cm}|}
\hline
\textbf{Question} & \textbf{Reply} & \textbf{Details} \\
\hline
\textbf{General:} & & \\
\hline
Does the system inform the user about the purpose of personalization?  & Yes & \\
\hline
Does the system inform the user who developed the technology and is liable in cases of wrongdoing?   & Yes & \\
\hline
Does the  system inform  the user about their rights under data protection law? & Partial & Local law rights are not specified. \\
\hline
Does the system inform the user about possible risks of engaging with the system? & No & Risks are not specified.\\
\hline
\textbf{Input:} & & \\
\hline
Have users given informed consent about the collection, processing, and storage of their data? & Partial & Default data collection policies are not specified. \\
\hline
Does the system inform the user about the fact that data is collected for personalization? & Yes & \\
\hline
Does the system inform the user about which data is collected to produce personalized content for them?  & Yes &  \\
\hline
Does the system inform the user about pre-processing done with the data collected for personalization purposes? & No & Pre-processing of data is not explained. \\
\hline
Does the system inform the user if their data is used and shared beyond the goals of personalization?   & Partial & Information about sharing data with partners is given without sufficient details as to the use of this data. \\
\hline
\textbf{Processing:} & & \\
\hline
Does the system inform the user about the kind of data that is processed to create a certain personalized item?  & Partial & The link between data sources and personalization is not clear. \\
\hline
Does the system explain to the user why they are receiving a certain personalization? & Partial & The notion of personalization is generally mentioned but not specified enough. \\
\hline
Does the system inform the user about the behavioural models underlying the personalization system? & No & Missing information about models used. \\
\hline
Does the system inform the user about possible constraints of the model such that may result from pre-processing or biases in the dataset? & No & Missing information about  models constraints. \\
\hline
\textbf{{}Output:} & & \\
\hline
Does the system present information to the user in a location where they can notice it and access it easily? & Partial & Hard to find the links to this data. Visibility and accessibility are lacking. \\
\hline
Does the system provide information to the user in a comprehensible way and can they act upon this information?  & Partial & Setting option is hard to understand and follow. \\
\hline
Does the system provide the user with information in a clear and simple language that avoids technical terms? & Yes &  \\
\hline
Does the system make it clear to the user that they interact with a machine? & Yes & \\
\hline
\textbf{Control:} & & \\
\hline
Does the system provide the user with the opportunity to specify their goals which are then used for personalization? & No & Missing capability. \\
\hline
Does the system provide the user with different options as to the personalized content they receive?  & Partial & Notification control is good. Ads control is poor. Data control is very partial. \\
\hline
Does the system provide the user with opt-in and opt-out options (e.g. for data collection)? & Partial & Complicated. Users have to control each option in separation. \\
\hline
If applicable, can the user adjust frequency and timing of personalized content? & Partial. & Is not supported for some content. \\
\hline
Does the user have a say in which data or models are used for personalization? & Partial & Users cannot fully understand the connection between data and personalization. \\
\hline
Does the system encourage the user to give feedback and express their opinion about the personalization mechanisms used (type, frequency, duration, etc.)? & No & Feedback is not strongly encouraged. \\
\hline
 \end{tabular}}
  \caption{Preliminary Checklist Application to Facebook}
  \label{tab:app}
\end{table*}

\begin{figure*} [t]
\centering
\includegraphics[width=13cm]{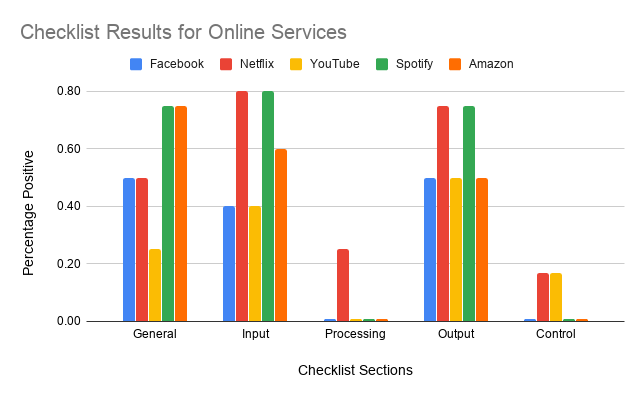}
\caption{Preliminary checklist, online sites: Y-axis is the percentage of positive replies in each checklist section}
\label{fig:comp}
\end{figure*}

To perform a preliminary comparison between the different sites and between the different sections of the checklist for each site, we also compute the percentage of \quotes{Yes} replies (i.e. full adherence to the best practices) for each checklist section. Namely, for each checklist question, we give a \quotes{Yes} reply a value of 1. We then sum these values for each section and divide it by the total number of questions in the corresponding section. This computation, while being limited and potentially biased due to the subjective filling of the checklist by the research team, may offer comparative information about the different sites and between the different checklist sections as to adequate coverage of transparency items. 
Figure~\ref{fig:comp} presents the result of this comparison.  
We further discuss these results in the next section. 


\section{Discussion}
\label{section:discuss}
\subsection{Advantages of the transparency checklist}
The major advantage of the transparency checklist is that it helps system designers understand where they are strong on transparency and where improvements are needed. Looking at Figure~\ref{fig:comp} and the online systems we have examined from the designer perspective, we notice that they primarily focus on realizing transparency in the \quotes{Input} category, i.e. with regard to data collection and the handling of user data. They are particularly weak in providing information about why and how models bring about certain personalization (\quotes{Processing}). They also lack participatory elements such as offering the user different options of personalization or allowing the user to supply feedback to the system (\quotes{Control}). 

This trend to follow best practices of data or \quotes{Input} transparency may be attributable to the rise of data protection laws such as the GDPR. System designers so far pay less attention to transparency about the reasoning and underlying logic of personalization. This is a severe shortcoming as ethics and philosophical work on transparency in algorithmic system clearly identify the need to disclose information about how a certain outcome (personalization) emerged. Making processing-related information transparent does not necessarily mean cracking open and looking inside the system, but rather providing meaningful and understandable information about the goals of personalization as well as the factors that contribute to making a personalized recommendation. 

We suspect that transparency about the reasoning of a system will gain relevance in the future. In fact, there is an ongoing debate whether the GDPR even provides a legal right to receive an explanation for algorithmic decision-making~\cite{Wachter.2016}. Legal cases in the future will shed light on such questions and eventually, disclosure of why and how a computer model caused a certain outcome may become customary practice. 

Literature also points to the need for user control to fulfill transparency~\cite{Simon.2010}. Users should be provided with different options of personalization to best align with their personal goals and increase their autonomy. Our application of the checklist points to significant shortcomings in the realm of \quotes{user control}. While both the areas of processing and user control exhibit a lack of transparency, increasing only one of the two areas would miss the point of transparency's comprehensive nature.  Disclosing the factors weighted in a personalized recommendation AND providing the user with the opportunity to adjust these factors meets the demands of transparency's comprehensive approach and potentially leads to more meaningful interaction between the user and the system.

As a system designer, having applied the checklist and seen some blind spots, one would now be able to make a deliberate decision about whether to increase transparency and user control in one's own system. 

\subsection{Transparency from multiple stakeholders' point of view}

Another advantage of the checklist is that it can be used as an assessment tool, not just internally for self-assessment but also as an openly accessible evaluation of a system's transparency performance. An online service may commission a \quotes{transparency check} by an independent organization to assess the system's trustworthiness. This application of the checklist may increase user's trust in a system. Studies show that transparency can be a competitive advantage of companies~\cite{Buell.2019}, and thus companies may have an interest in providing information about the transparency levels of their online services. The desire for independent audits may increase in the future with movements to certify \quotes{trustworthy use of Artificial Intelligence}~\cite{FraunhoferInstituteforIntelligentAnalysisandInformationSystems.2019}. 

Beyond commissioned reviews of a system's transparency performance, users and activists may employ the transparency checklist as a means of control and oversight of online services. A comparison of online services' transparency performance as in figure 1 exposes the brands behind them and may generate pressure to implement increased transparency. The checklist can thus be used as a means to raise awareness of algorithmic transparency, and may be adopted by like-minded institutions and projects (e.g. CyCat\footnote{\url{http://www.cycat.io/}}).

We should note here that, while an ethics perspective promotes user control and meaningful transparency, it is not certain that end users desire transparency and control. From extensive work in the field of privacy and data protection, we know that users claim privacy to be an important issue for them but rarely take steps to protect their data (“privacy paradox”)~\cite{Gerber.2018}. Similar dynamics may apply to transparency. Therefore, feedback from end users on how much transparency and what kind of transparency they prioritize in a system would be helpful for system designers. Such feedback could be obtained in further studies or in collaborative designs with the active involvement of end users. Nevertheless, independent of users’ personal preferences, users should have the opportunity to take advantage of transparency. Even when users disregard information provided to them, system designers have an ethical responsibility to implement transparency best practices.

Another significant issue concerns the relationship of information to the user. Transparency is a relational concept. The same information may make something transparent to one group or individual but not to others~\cite{Vakarelov.2019}. It follows that transparency must be configured to the individual user. In fact, we may need personalization technology to fulfill the transparency best practices we have suggested in this paper~\cite{Kouki.2019, schneider2019personalized}.

\subsection{Limitations of transparency}

We now briefly point to some limitations of our approach. We note that the idea or \quotes{ideal} of transparency itself has limits~\cite{Ananny.2018}. For instance, transparency rests on the idea that something can be known~\cite{Ananny.2018}. There is no guarantee that we succeed in understanding a model, even when transparency is in place. This can be due to lack of resources, human capital~\cite{Mittelstadt.2016}, and lack of basic digital or technical literacy~\cite{Burrell.2016}. Disclosing information can also confuse users and not add to clarity or insight~\cite{Bannister.2011, Ananny.2018}. Transparency may further clash with important ethical principles such as privacy. Full disclosure of input or output data may put users at risk of being re-identified, especially in areas like finance and medicine~\cite{Lepri.2018}. Business interests may also be legitimate reasons to reject full disclosure~\cite{Lepri.2018}. 

These limitations of transparency also put a checklist in perspective. Whether transparency is appropriate or warranted depends on the unique use case. It remains an open question how much and what kind of transparency should be provided. These are questions for the personalization community or the respective design teams. Ideally, such questions will be answered in collaboration with end users. 


\section{Conclusion and Future Work}
\label{section:conclude}
In this work, we have presented our best practices for transparency in Aritifical Intelligence based personalization systems and we have applied our own transparency checklist to existing systems. While transparency needs may vary depending on the use case, the checklist is valuable as a supporting instrument that guides system designers in embedding transparency into their work. We have demonstrated such a use by a preliminary application of the checklist from a system designer perpective to prominent AI based services that use personalization. 

While we propose a first transparency and user control checklist, we recognize that it may be amended in future engagement with researchers and system designers. An important next step is to have an exchange with practitioners in the field and develop a consensus regarding a transparency checklist for the personalization community~\cite{aczel2019consensus}. This can increase the checklist’s likelihood of adoption. We therefore encourage researchers, funding agencies, and journals to provide feedback and recommendations. Furthermore, tangible design actions based on the best practices have to be developed in future work. Tutorials and workshops may invite system designers to apply the checklist and create innovative design solutions that implement transparency in their respective systems. 

Another line of research that follows up on this work relates to end users' transparency needs. On the one hand, studies may provide additional insights into how transparency helps or hinders end users in their engagement with a system. On the other hand, further research is required to understand the respective transparency needs of diverse end users. Possibilities to personalize transparency should be explored to ensure that users of different capabilities receive tailored information and user control options.

\ack 
This project has received funding from the European Union's Horizon 2020 WeNet research and innovation program under grant agreement No 823783.
The authors kindly thank participants of the ACM UMAP 2020 conference session \quotes{Demo and Late-Breaking Results} for their comments and questions on an earlier version of this paper.
We further thank our colleagues PD Dr. Jessica Heesen and Michael Lohaus for their feedback on our research. 

\bibliography{transRef}

\begin{thebibliography}{10}

\bibitem{link_uxdesign}
\url{https://uxdesign.cc/user-experience-vs-business-goals-finding-the-balance-7507ac85b0a9},
  2019.

\bibitem{InstituteofElectricalandElectronicsEngineers.2019}
Ethically aligned design: A vision for prioritizing human well-being with
  autonomous and intelligent systems, 2019.

\bibitem{link_latlng_03}
\url{https://www.brookings.edu/research/the-case-for-ai-transparency-requirements/},
  2020.

\bibitem{EuropeanParliamentandCounciloftheEuropeanUnion.27April2016}
General data protection~regulation (regulation (eu) 2016/679), 27 April 2016.

\bibitem{Abdollahi.C2018}
Behnoush Abdollahi and Olfa Nasraoui, `Transparency in fair machine learning:
  the case of explainable recommender systems', in {\em Human and Machine
  Learning}, eds., Jianlong Zhou and Fang Chen, volume~44 of {\em
  Human-computer interaction series},  21--35, Springer, Cham, Switzerland, (C
  2018).

\bibitem{ACMCouncil.22June2018}
{ACM Council}.
\newblock Acm code of ethics and professional conduct, 2018.

\bibitem{aczel2019consensus}
Balazs Aczel, Barnabas Szaszi, Alexandra Sarafoglou, Zoltan Kekecs, {\v{S}}imon
  Kucharsk{\`y}, Daniel Benjamin, Christopher~D Chambers, Agneta Fisher, Andrew
  Gelman, Morton~A Gernsbacher, et~al., `A consensus-based transparency
  checklist', {\em Nature human behaviour},  1--3, (2019).

\bibitem{adadi2018peeking}
Amina Adadi and Mohammed Berrada, `Peeking inside the black-box: A survey on
  explainable artificial intelligence (xai)', {\em IEEE Access}, {\bf 6},
  52138--52160, (2018).

\bibitem{AIEthicsImpactGroup.2020}
{AI Ethics Impact Group}.
\newblock From principles to practice: An interdisciplinary framework to
  operationalise ai ethics.

\bibitem{Amershi.2019}
Saleema Amershi, Kori Inkpen, Jaime Teevan, Ruth Kikin-Gil, Eric Horvitz, Dan
  Weld, Mihaela Vorvoreanu, Adam Fourney, Besmira Nushi, Penny Collisson, Jina
  Suh, Shamsi Iqbal, and Paul~N. Bennett, `Guidelines for human-ai
  interaction', in {\em Proceedings of the 2019 CHI Conference on Human Factors
  in Computing Systems - CHI '19}, eds., Stephen Brewster, Geraldine
  Fitzpatrick, Anna Cox, and Vassilis Kostakos, pp. 1--13, New York, New York,
  USA, (2019). {ACM Press}.

\bibitem{Ananny.2018}
Mike Ananny and Kate Crawford, `Seeing without knowing: Limitations of the
  transparency ideal and its application to algorithmic accountability', {\em
  New Media {\&} Society}, {\bf 20}(3),  973--989, (2018).

\bibitem{Bannister.2011}
Frank Bannister and Regina Connolly, `The trouble with transparency: A critical
  review of openness in e-government', {\em Policy {\&} Internet}, {\bf 3}(1),
  158--187, (2011).

\bibitem{Benkler.2006}
Yochai Benkler, {\em The Wealth of Networks: How Social Production Transforms
  Markets and Freedom}, {Yale University Press}, New Haven and London, 2006.

\bibitem{boell2012conceptualizing}
Sebastian Boell and Dubravka Cecez-Kecmanov, `Conceptualizing information
  systems: From'input-processing-output'devices to sociomaterial apparatuses',
  (2012).

\bibitem{borji2012state}
Ali Borji and Laurent Itti, `State-of-the-art in visual attention modeling',
  {\em IEEE transactions on pattern analysis and machine intelligence}, {\bf
  35}(1),  185--207, (2012).

\bibitem{Bozdag.2013}
Engin Bozdag, `Bias in algorithmic filtering and personalization', {\em Ethics
  and Information Technology}, {\bf 15}(3),  209--227, (2013).

\bibitem{Brill.2015}
J~Brill, `Scalable approaches to transparency and accountability in
  decisionmaking algorithms: remarks at the nyu conference on algorithms and
  accountability', {\em Federal Trade Commission}, {\bf 28}, (2015).

\bibitem{Buell.2019}
Ryan~W. Buell and MoonSoo Choi, `Improving customer compatibility with
  operational transparency', {\em SSRN Electronic Journal}, (2019).

\bibitem{Buolamwini.2018}
Joy Buolamwini and Timnit Gebru, `Gender shades: Intersectional accuracy
  disparities in commercial gender classification', {\em Proceedings of Machine
  Learning Research}, {\bf 81},  1--15, (2018).

\bibitem{Burrell.2016}
Jenna Burrell, `How the machine `thinks': Understanding opacity in machine
  learning algorithms', {\em Big Data {\&} Society}, {\bf 3}(1),
  205395171562251, (2016).

\bibitem{Chromnik.20March2019}
Michael Chromnik, Malin Eiband, Sarah~Theres V{\"o}lk, and Daniel Buschek.
\newblock Dark patterns of explainability, transparency, and user controlfor
  intelligent systems, 20 March 2019.

\bibitem{Crain.2019}
Crain and Nadler, `Political manipulation and internet advertising
  infrastructure', {\em Journal of Information Policy}, {\bf 9},  370, (2019).

\bibitem{Dahl.2018}
E.~S. Dahl, `Appraising black-boxed technology: the positive prospects', {\em
  Philosophy {\&} Technology}, {\bf 31}(4),  571--591, (2018).

\bibitem{Laat.2018}
Paul~B. de~Laat, `Algorithmic decision-making based on machine learning from
  big data: Can transparency restore accountability?', {\em Philosophy {\&}
  Technology}, {\bf 31}(4),  525--541, (2018).

\bibitem{deci2002overview}
Edward~L Deci and Richard~M Ryan, `Overview of self-determination theory: An
  organismic dialectical perspective', {\em Handbook of self-determination
  research},  3--33, (2002).

\bibitem{Floridi.2018}
Luciano Floridi, Josh Cowls, Monica Beltrametti, Raja Chatila, Patrice
  Chazerand, Virginia Dignum, Christoph Luetge, Robert Madelin, Ugo Pagallo,
  Francesca Rossi, Burkhard Schafer, Peggy Valcke, and Effy Vayena,
  `Ai4people---an ethical framework for a good ai society: Opportunities,
  risks, principles, and recommendations', {\em Minds and Machines}, {\bf
  28}(4),  689--707, (2018).

\bibitem{FraunhoferInstituteforIntelligentAnalysisandInformationSystems.2019}
{Fraunhofer Institute for Intelligent Analysis and Information Systems}.
\newblock Trustworthy use of artificial intelligence.

\bibitem{Gerber.2018}
Nina Gerber, Paul Gerber, and Melanie Volkamer, `Explaining the privacy
  paradox: A systematic review of literature investigating privacy attitude and
  behavior', {\em Computers \& Security}, {\bf 77},  226--261, (2018).

\bibitem{harambam2019designing}
Jaron Harambam, Dimitrios Bountouridis, Mykola Makhortykh, and Joris
  Van~Hoboken, `Designing for the better by taking users into account: a
  qualitative evaluation of user control mechanisms in (news) recommender
  systems', in {\em Proceedings of the 13th ACM Conference on Recommender
  Systems}, pp. 69--77, (2019).

\bibitem{hartford2016deep}
Jason~S Hartford, James~R Wright, and Kevin Leyton-Brown, `Deep learning for
  predicting human strategic behavior', in {\em Advances in Neural Information
  Processing Systems}, pp. 2424--2432, (2016).

\bibitem{heer2019agency}
Jeffrey Heer, `Agency plus automation: Designing artificial intelligence into
  interactive systems', {\em Proceedings of the National Academy of Sciences},
  {\bf 116}(6),  1844--1850, (2019).

\bibitem{HighLevelExpertGrouponArtificialIntelligence.2019}
{High-Level Expert Group on Artificial Intelligence}.
\newblock Ethics guidelines for trustworthy ai.

\bibitem{unknown}
Robert Hoffman, Shane Mueller, Gary Klein, and Jordan Litman.
\newblock Metrics for explainable ai: Challenges and prospects, 12 2018.

\bibitem{Hois.2019}
Joana Hois, Dimitra Theofanou-Fuelbier, and Alischa~Janine Junk, `How to
  achieve explainability and transparency in human ai interaction', in {\em HCI
  INTERNATIONAL 2019 - POSTERS}, ed., Constantine Stephanidis, volume 1033 of
  {\em Communications in Computer and Information Science},  177--183,
  Springer, [Place of publication not identified], (2019).

\bibitem{jackson2016encouraging}
Corey Jackson, Gabriel Mugar, Kevin Crowston, and Carsten {\O}sterlund,
  `Encouraging work in citizen science: Experiments in goal setting and
  anchoring', in {\em Proceedings of the 19th ACM Conference on Computer
  Supported Cooperative Work and Social Computing Companion}, pp. 297--300,
  (2016).

\bibitem{Kitchin.2017}
Rob Kitchin, `Thinking critically about and researching algorithms', {\em
  Information, Communication {\&} Society}, {\bf 20}(1),  14--29, (2017).

\bibitem{kolekar2019rule}
Sucheta~V Kolekar, Radhika~M Pai, and Manohara~Pai MM, `Rule based adaptive
  user interface for adaptive e-learning system', {\em Education and
  Information Technologies}, {\bf 24}(1),  613--641, (2019).

\bibitem{Kouki.2019}
Pigi Kouki, James Schaffer, Jay Pujara, John O'Donovan, and Lise Getoor,
  `Personalized explanations for hybrid recommender systems', in {\em
  Proceedings of the 24th International Conference on Intelligent User
  Interfaces}, eds., Wai-Tat Fu, Shimei Pan, Oliver Brdiczka, Polo Chau, and
  Gaelle Calvary, pp. 379--390, New York, NY, USA, (2019). ACM.

\bibitem{Lanzing.2016}
Marjolein Lanzing, `The transparent self', {\em Ethics and Information
  Technology}, {\bf 18}(1),  9--16, (2016).

\bibitem{Lepri.2018}
Bruno Lepri, Nuria Oliver, Emmanuel Letouz{\'e}, Alex Pentland, and Patrick
  Vinck, `Fair, transparent, and accountable algorithmic decision-making
  processes', {\em Philosophy {\&} Technology}, {\bf 31}(4),  611--627, (2018).

\bibitem{Miller.22062017}
Tim Miller.
\newblock Explanation in artificial intelligence: Insights from the social
  sciences.

\bibitem{Mittelstadt.2016}
Brent~Daniel Mittelstadt, Patrick Allo, Mariarosaria Taddeo, Sandra Wachter,
  and Luciano Floridi, `The ethics of algorithms: Mapping the debate', {\em Big
  Data {\&} Society}, {\bf 3}(2),  205395171667967, (2016).

\bibitem{muldoon2018survey}
Conor Muldoon, Michael~J O’Grady, and Gregory~MP O’Hare, `A survey of
  incentive engineering for crowdsourcing', {\em The Knowledge Engineering
  Review}, {\bf 33}, (2018).

\bibitem{ONeil.2016}
Cathy O'Neil, {\em Weapons of Math Destruction: How Big Data Increases
  Inequality and Threatens Democracy}, {Allen Lane}, London, 2016.

\bibitem{Pariser.2012}
Eli Pariser, {\em The Filter Bubble: What the Internet is Hiding From You},
  {Penguin Books}, London, 2012.

\bibitem{Pasquale.2015}
Frank Pasquale, {\em The Black Box Society: The Secret Algorithms that Control
  Money and Information}, {Harvard University Press}, Cambridge, 2015.

\bibitem{plonsky2019predicting}
Ori Plonsky, Reut Apel, Eyal Ert, Moshe Tennenholtz, David Bourgin, Joshua~C
  Peterson, Daniel Reichman, Thomas~L Griffiths, Stuart~J Russell, Evan~C
  Carter, et~al., `Predicting human decisions with behavioral theories and
  machine learning', {\em arXiv preprint arXiv:1904.06866}, (2019).

\bibitem{Preece.2018}
Alun Preece, `Asking `why' in ai: Explainability of intelligent systems -
  perspectives and challenges', {\em Intelligent Systems in Accounting, Finance
  and Management}, {\bf 25}(2),  63--72, (2018).

\bibitem{Rosenfeld.2019}
Avi Rosenfeld and Ariella Richardson, `Explainability in human--agent systems',
  {\em Autonomous Agents and Multi-Agent Systems}, {\bf 33}(6),  673--705,
  (2019).

\bibitem{schneider2019personalized}
Johanes Schneider and Joshua Handali, `Personalized explanation in machine
  learning: A conceptualization', {\em arXiv preprint arXiv:1901.00770},
  (2019).

\bibitem{Seaver.2019}
Nick Seaver, `Knowing algorithms', in {\em DigitalSTS}, eds., Janet Vertesi and
  David Ribes, {Princeton University Press}, Princeton, New Jersey, (2019).

\bibitem{segal2018optimizing}
Avi Segal, Kobi Gal, Ece Kamar, Eric Horvitz, and Grant Miller, `Optimizing
  interventions via offline policy evaluation: studies in citizen science', in
  {\em Thirty-Second AAAI Conference on Artificial Intelligence}, (2018).

\bibitem{segal2016intervention}
Avi Segal, Ece Kamar, Eric Horvitz, Alex Bowyer, Grant Miller, et~al.,
  `Intervention strategies for increasing engagement in crowdsourcing:
  Platform, predictions, and experiments', (2016).

\bibitem{Simon.2010}
Judith Simon, `The entanglement of trust and knowledge on the web', {\em Ethics
  and Information Technology}, {\bf 12}(4),  343--355, (2010).

\bibitem{Srmo.2005}
Frode S{\o}rmo, J{\"o}rg Cassens, and Agnar Aamodt, `Explanation in case-based
  reasoning--perspectives and goals', {\em Artificial Intelligence Review},
  {\bf 24}(2),  109--143, (2005).

\bibitem{Spencer.2019}
Shaun~B. Spencer, `The problem of online manipulation', {\em SSRN Electronic
  Journal}, (2019).

\bibitem{Steiner.2013}
Christopher Steiner, {\em Automate This: How Algorithms Came to Rule Our
  World}, Portfolio/Penguin, New York, N.Y., 2013.

\bibitem{Susser.2018}
Daniel Susser, Beate Roessler, and Helen~F. Nissenbaum, `Online manipulation:
  Hidden influences in a digital world', {\em SSRN Electronic Journal}, (2018).

\bibitem{Tene.2013}
Omer Tene and Jules Polonetsky, `Big data for all: Privacy and user control in
  the age of analytics', {\em Northwestern Journal of Technology and
  Intellectual Property}, {\bf 239}, (2013).

\bibitem{Turilli.2009}
Matteo Turilli and Luciano Floridi, `The ethics of information transparency',
  {\em Ethics and Information Technology}, {\bf 11}(2),  105--112, (2009).

\bibitem{Vakarelov.2019}
Orlin Vakarelov and Kenneth Rogerson, `The transparency game: Government
  information, access, and actionability', {\em Philosophy {\&} Technology},
  {\bf 23}(1--2),  193, (2019).

\bibitem{van2019collecting}
Rob van Roy, Sebastian Deterding, and Bieke Zaman, `Collecting pok{\'e}mon or
  receiving rewards? how people functionalise badges in gamified online
  learning environments in the wild', {\em International Journal of
  Human-Computer Studies}, {\bf 127},  62--80, (2019).

\bibitem{Veltri.2019}
Giuseppe~A. Veltri, `The political bubble', {\em Longitude}, {\bf 93},  54--59,
  (2019).

\bibitem{Wachter.2016}
Sandra Wachter, Brent Mittelstadt, and Luciano Floridi, `Why a right to
  explanation of automated decision-making does not exist in the general data
  protection regulation', {\em SSRN Electronic Journal}, (2016).

\bibitem{yanovsky2019one}
Stav Yanovsky, Nicholas Hoernle, Omer Lev, and Kobi Gal, `One size does not fit
  all: Badge behavior in q\&a sites', in {\em Proceedings of the 27th ACM
  Conference on User Modeling, Adaptation and Personalization}, pp. 113--120,
  (2019).

\bibitem{Zerilli.2019}
John Zerilli, Alistair Knott, James Maclaurin, and Colin Gavaghan,
  `Transparency in algorithmic and human decision-making: Is there a double
  standard?', {\em Philosophy {\&} Technology}, {\bf 32}(4),  661--683, (2019).

\bibitem{Zhou.C2018}
{\em Human and Machine Learning: Visible, Explainable, Trustworthy and
  Transparent}, eds., Jianlong Zhou and Fang Chen, Human-computer interaction
  series, Springer, Cham, Switzerland, C 2018.

\bibitem{Zou.2018}
James Zou and Londa Schiebinger, `Ai can be sexist and racist - it's time to
  make it fair', {\em Nature}, {\bf 559}(7714),  324--326, (2018).

\end{thebibliography}
\end{document}